\documentclass{optica-article}

\journal{opticajournal} % for journals or Optica Open

\articletype{Research Article}

\usepackage[separate-uncertainty=true]{siunitx}
\usepackage{lineno}
\begin{document}

\title{Hermite-Gaussian multi-mode excitation in spatial gain shaping approaches for laser resonators}

\author{Michael Zwilich\authormark{*} and Carsten Fallnich}

\address{University of Münster, Institute of Applied Physics, Corrensstr. 2, 48149 Münster}

\email{\authormark{*}m.zwilich@uni-muenster.de}

\begin{abstract*}
This study explores the excitation of transverse laser modes through spatial gain shaping, while focusing on the boundary between selective single-mode and multi-mode lasing. By deliberately reducing the similarity between intensity distributions of pump and laser mode, it was studied, if and which other modes are excited besides the target mode, and how the modes compete for the spatially distributed gain. Analysis of the usually unwanted multi-mode lasing revealed characteristic properties of pump distributions adapted to Hermite-Gaussian $\text{HG}_{m,0}$ modes: a center-heavy pump distribution at first distinctly excites the target mode and eventually low-order modes, whereas an eccentric pump distribution reduces the lasing threshold at the expense of distinction to the neighboring modes. By understanding why certain gain distributions do not excite a single mode, we infer guidelines for the design of pump patterns in spatial gain shaping approaches.
\end{abstract*}

%%%%%%%%%%%%%%%%%%%%%%%%%%  body  %%%%%%%%%%%%%%%%%%%%%%%%%%
\section{Introduction}
% Importance and applications of transverse (resonator) modes
Laser light can be formed into diverse spatial patterns~\cite{Forbes2019} and the so-called 'structured light' shows intriguing properties, which enable advances in optical trapping~\cite{Chen2013}, communication schemes~\cite{Willner2015} and a multitude of other research areas~\cite{Ast2021, Heinze2022}.
The Hermite- (HG) and Laguerre-Gaussian (LG) mode families serve as fundamental components in the generation and description of spatially structured light, because they provide a complete basis for expressing arbitrary fields via their superposition.
Light distributions of HG and LG modes can be generated by reshaping an existing laser beam~\cite{Beijersbergen1993, Matsumoto2008, Uren2019} or they can be realized directly from a laser resonator, because the eigenmodes of free-space resonators are well approximated by HG and LG modes~\cite{Boyd1962, Siegman1986}.

% Methods of transverse mode generation
To directly excite a transverse mode of a laser resonator, spatially structured loss elements adapted to the desired mode profile, e.g., wires~\cite{Chu2012} or printed masks~\cite{Lukowski2017}, can be introduced, such that laser oscillation is limited to a single target mode.
The inflexibility of such physical loss elements is overcome by programmable intra-cavity phase shaping. 
For instance, using a liquid-crystal based spatial light modulator in a laser cavity allows to produce a variety of beams with different transverse intensity distributions~\cite{Ngcobo2013}.
When intra-cavity elements are not tolerable due to their losses or the resonator is not accessible, such as with microchip lasers~\cite{Kong2012}, modes can also be selectively excited by providing spatially structured gain matching the intensity distribution of a target mode.

% Spatial gain shaping
In its simplest form, spatial gain shaping can be realized by focusing a pump beam off-axis, exploiting the different spatial intensity distributions of the transverse modes~\cite{Laabs1996, Litvin2017, Lukowski2019}.
These approaches are easy to implement, but the realized gain distributions are rather mode-unspecific, meaning they are only matched to parts of the target mode, e.g., the outermost lobe of a HG mode.
Off-axis pumping reportedly faces difficulties in selectively exciting higher-order modes, because with increasing mode-order the outermost lobes change only little in their position, so that adjacent modes with similar intensity distributions experience sufficient gain and are excited as well~\cite{Chen1997}.
Additionally, excitation of two-dimensional $\text{HG}_{m,n}$ modes usually requires resonator modifications~\cite{Pan2020, Sheng2024}, because one-dimensional $\text{HG}_{m,0}$ modes are strongly favored through a greater overlap between gain and laser mode distributions, even when the pump beam is displaced along both transverse dimensions due to a lack of defined modal axes in a cylindrically symmetric resonator.

% Advanced shaping with spatial light modulators
With spatial light modulators the gain distribution can be better matched to a target mode by making use of the mode's complete intensity distribution.
For instance, a diffractive optical element (DOE) allows to reshape pump light into a different form, e.g., LG mode-like ring-shaped intensity distributions can be realized with an axicon~\cite{Zhang2021}.
While each manufactured DOE is limited to a single realizable pump distribution, a digital micromirror device (DMD) acting as a programmable, and therefore flexible, spatial light modulator allows to more precisely match the pump beam to the intensity distribution of a target mode, evidenced by selective excitation of nearly \num{1000} $\text{HG}_{m,n}$ modes~\cite{Schepers2019}.
By displaying a two-dimensional binary image on the DMD, which encodes the reflection from the individually addressable micromirrors, a pump beam with the desired spatial intensity distribution is formed in reflection.

While all of these approaches present successful excitation of single transverse modes, there is a lack of systematic investigations on the boundary between selective single-mode and multi-mode operation.
Here, we specifically explore this regime of simultaneous excitation of multiple transverse modes to gain insights about their practical interactions in laser resonators.
By superposing two pump beams, one shaped by a DMD specifically adapted to a target mode and the other mode-unspecific, pump distributions with varying degrees of similarity to a target mode are generated.
This allows to study, how robust pump distributions are to additionally supplied gain, if and which other modes are excited besides the target mode, and how the modes compete for the spatially distributed gain.
By understanding why certain gain distributions do not excite a single mode, we infer guidelines for the design of pump patterns in spatial gain shaping approaches.

\section{Experimental setup} \label{sec:setup}
\begin{figure}[!ht]
	\centering
	\includegraphics{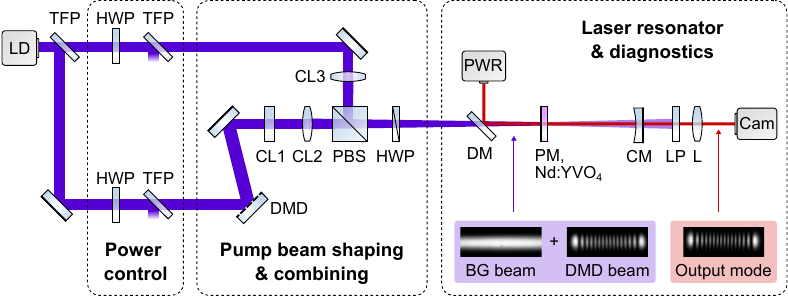}
	\caption{Schematic of the experimental setup. One pump beam was shaped according to the intensity distribution of a target mode with a DMD. A second pump beam (BG, background beam) was shaped mode-unspecifically by a cylindrical lens. For details, see text.}
	\label{fig:setup}
\end{figure}

In spatial gain shaping, the pump distribution is adapted to the intensity profile of a target mode.
Here, to study how strongly a pump distribution has to be adapted to a target mode for excitation of only that mode, two pump beams were shaped with varying degrees of mode-specificity: one mode-specific beam (DMD beam) similar to the intensity profile of a target mode and one mode-unspecific 'background' beam (BG beam) without adaption to any mode.
These two pump beams were superposed in the active medium of a subsequent laser resonator and the implications on transverse mode excitation are investigated (see Fig.~\ref{fig:setup} for the experimental setup). 

% Resonator
A $\qty{6}{\mm} \times \qty{6}{\mm} \times \qty{1}{\mm}$ a-cut Nd:YVO$_4$ crystal with 1~at.\si{\percent} doping concentration was used as the laser medium. 
It was wrapped in indium foil and mounted in a copper heat sink, which was water-cooled to room temperature.
The crystal's outward facing side served as a plane mirror (PM) in a plano-concave resonator, as it was dielectrically coated for transmission of the pump wavelength \qty{808}{\nm} and for reflection of the laser wavelength \qty{1064}{\nm} with a residual transmission of \qty{0.1}{\percent}.
The crystal's inward facing side was anti-reflection coated for pump and laser wavelength. 
The resonator was closed by a curved mirror (CM) with a radius of curvature $R=\qty{-1}{\m}$ and a transmission of \qty{0.5}{\percent} at the laser wavelength.
The laser output power was measured by a power meter (PWR) via the low-power output through the crystal, which was separated from the pump beam by a dichroic mirror (DM).
To evaluate the spatial intensity profile of the output beam, the crystal plane was imaged onto a camera (Cam) with a lens (L) through the curved resonator end mirror, using a long-pass filter (LP) for blocking residual pump light.

The crystal was pumped by light emitted from a fiber-coupled laser diode (LD) at a wavelength of \qty{808}{\nm}.
This initial pump beam was split by a thin-film polarizer (TFP) into orthogonally polarized components of equal power, which were directed along two distinct optical paths, where they were spatially modified and controlled in their powers, and recombined afterwards at a polarizing beamsplitter (PBS) for optical pumping of the gain medium.

% Mode specific spatial gain shaping by DMD
The p-polarized pump beam component was spatially modulated with a DMD (an array of $912\times 1140$ micromirrors, each with a side length of \qty{7.6}{\um}, Texas Instruments DLP4500NIR~\cite{TexasInstruments2022}) by tilting each of its micromirrors into an 'on'- or 'off'-state. 
A spatial gain distribution adapted to a specific transverse resonator mode was realized by displaying an according binary image on the DMD, thereby shaping the reflected pump beam distribution (see Sec.~\ref{subsec:dmd_shaping} for a detailed description of the mode-specific beam shaping).
The DMD was oriented such that the incidence angle and the tilting angle of the micromirrors fulfilled the Blaze condition, concentrating most of the reflected power (\qty{76}{\percent}) in one diffraction order~\cite{Rice2009, Deng2022}.
The power incident on the DMD was limited to \qty{7}{\W} based on the recommended operating conditions given by the manufacturer~\cite{TexasInstruments2022}.
To still obtain maximum possible gain with this power-limited beam, its absorption in the laser medium was maximized by matching its direction of polarization to a crystal axis using a half-wave plate (HWP).

% Imaging vs focusing
For these first experiments, only $\text{HG}_{m,0}$ modes were excited, i.e., modes with a Gaussian intensity evolution along the vertical direction and an arrangement of $m+1$ nearly-equidistant peaks along the horizontal direction.
Consequently, to match the pump beam in shape and size to the resonator mode (beam radius $w_0=\qty{381}{\um}$), along the horizontal direction the DMD plane was imaged (magnification ratio $M=\num{0.9}$) onto the crystal plane using a cylindrical lens (CL1, $f=\qty{150}{\mm}$). 
In contrast, along the vertical direction the DMD pump beam was focused with a second, orthogonally oriented, cylindrical lens (CL2, $f=\qty{100}{\mm}$) down to a vertical beam radius of $\qty{340}{\um}$, thus ensuring $\text{HG}_{m,0}$ modes by only exciting the vertical fundamental mode order $n=0$.

% Background beam
The s-polarized pump beam component provided mode-unspecific gain and is referred to as background beam. 
It was focused by a cylindrical lens (CL3, $f=\qty{100}{\mm}$), such that the beam was strongly astigmatic in the crystal plane, with beam radii of \qty{4.9}{\mm} in the horizontal and \qty{346}{\um} in the vertical direction. 
Therefore, the background beam had overlap with a multitude of transverse modes in the horizontal direction because of its great width compared to the resonator's fundamental transverse mode, whereas it only provided sufficient gain for the fundamental mode in the vertical direction.
By superposing the mode-specific DMD beam and the mode-unspecific background beam effective pump distributions with varying degrees of similarity to a target mode were generated (see Sec.~\ref{subsec:thresholds}).
How similar the effective pump distribution was to a target mode could be controlled by the power ratio of the two pump beams, whose powers were independently regulated by a HWP and a TFP in each of the optical paths ('Power control' in Fig.~\ref{fig:setup}) prior to being shaped by the DMD and CL3, respectively.

\section{Pump patterns and lasing thresholds}\label{sec:patterns_thresholds}
\subsection{Non-binary pump beam shaping with a DMD} \label{subsec:dmd_shaping}
\begin{figure}[!ht]
	\centering
	\includegraphics{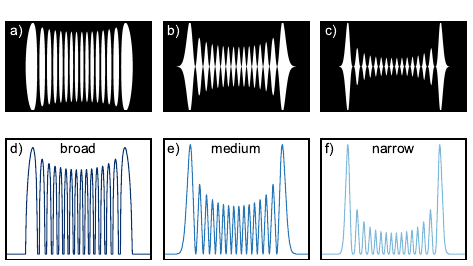}
	\caption{a), b), c) Exemplary binary patterns displayed on the DMD for excitation of the $\text{HG}_{15,0}$ mode. Light incident on white areas was reflected in the direction of the gain medium. d), e), f) Corresponding intensity distributions after projection onto the horizontal axis, analogous to the focusing of the beam along the vertical direction in the experiment.}
	\label{fig:pump_patterns}
\end{figure}

Despite the binary DMD architecture, gray scale intensity distributions can be realized, either with a superpixel approach~\cite{Ren2015}, which decreases the effective spatial resolution, or by rapidly flipping the micromirrors with varying duty-cycles~\cite{Graff2013}, which results in pump power variations over time.
To ensure a temporally stable pump pattern with high spatial resolution, an alternative approach was used, in which the vertical dimension of the DMD was turned into an intensity axis, leveraging the 'quasi one-dimensionality' of the $\textrm{HG}_{m,0}$ modes.
Along the horizontal direction of the DMD a binary pattern was displayed which matched the horizontal spatial features of a $\textrm{HG}_{m,0}$ mode, i.e., the characteristic sequence of $m+1$ peaks.
In contrast, the vertical DMD dimension was used to realize a non-binary power value at each horizontal position: graphically speaking, the number of 'on'-pixels in each column (depicted in white in Fig.~\ref{fig:pump_patterns}) controlled the intensity at each horizontal position.
As described in the experimental setup, the pump beam shaped by this spatially modulated reflection by the DMD was focused with a cylindrical lens along the vertical direction, essentially integrating the pump power along this dimension, so that at each horizontal position the vertical spatial dependence vanished and only the desired intensity remained.
The intensity could be controlled in increments equal to the number of micromirror rows on the DMD, which for the used device was \num{570} (equivalent to a resolution of about \num{9} bit), when assuming homogeneous illumination of the DMD.

% Mode-like pumping
In order to excite a certain transverse mode through pump beam shaping, the binary pattern displayed on the DMD was derived from the intensity distribution of that target mode.
For each mode three differently shaped patterns were investigated (see Fig.~\ref{fig:pump_patterns} for the patterns to excite $\text{HG}_{15,0}$).
% Pump patterns
The 'medium' pump pattern was designed to result in a pump distribution equal to the intensity of the target mode by making use of the above outlined shaping approach: the desired mode profile was displayed along the horizontal axis, whereas the vertical axis represented the corresponding intensities (see Fig.~\ref{fig:pump_patterns}b).
When focusing a so shaped pump beam along the vertical direction, the intensity distribution of the target mode was obtained (see Fig.~\ref{fig:pump_patterns}e).
The 'narrow' pattern corresponded to the square of the target mode's intensity, leading to more narrow spatial features and decreased power in the center of the pump beam.
In contrast, for the 'broad' pump pattern the two-dimensional intensity profile of the target mode was first stretched along the vertical direction to maximize reflected power, normalized to peak intensity and then converted to a binary mask by applying a simple threshold $\eta=\num{0.11}$, i.e., all pixels with intensities greater than $\eta$ were set to 'on' and all other pixels were turned 'off'~\cite{Schepers2019}.
% Shaping efficiency decreases
The pump patterns become narrower as more and more micromirrors are positioned in their 'off'-state (black areas in Fig.~\ref{fig:pump_patterns}), so that the reflected light power decreases from \qty{37}{\percent} (broad) to \qty{24}{\percent} (medium) and \qty{12}{\percent} (narrow) for the shown $\text{HG}_{15,0}$ patterns compared to all micromirrors switched 'on', i.e., \qty{100}{\percent} pump power.

\subsection{Characteristic lasing thresholds} \label{subsec:thresholds}
% Gain shaping for selective mode excitation
Laser oscillation in a single transverse mode requires that one mode is provided with enough gain to equal the resonator losses, while no other mode is excited along with it. 
The pump power necessary to excite a transverse mode is termed the lasing threshold $P_\text{th}$ and it is proportional to the effective mode volume
\begin{equation}
	V_\text{eff} = \left(\iiint I_\text{m,n}(x,y,z) \, I_\text{pump}(x,y,z) \, \text{d}V\right)^{-1}, \label{eq:modevolume}
\end{equation}
which is given by the spatial overlap between the normalized photon density distributions of laser mode $I_\text{m,n}$ and pump beam $I_\text{pump}$ over the volume $V$ of the gain medium~\cite{Kubodera1979}.
Thus, to preferably excite a transverse mode, the pump distribution should be mode-specific, i.e., adapted to the desired mode profile.
Ideally the pump distribution is shaped, such that the lasing threshold of the target mode is not only the lowest but ideally also well distinguished from the thresholds of all the other modes, favoring excitation of the target mode only.

\begin{figure}[!ht]
	\centering
	\includegraphics{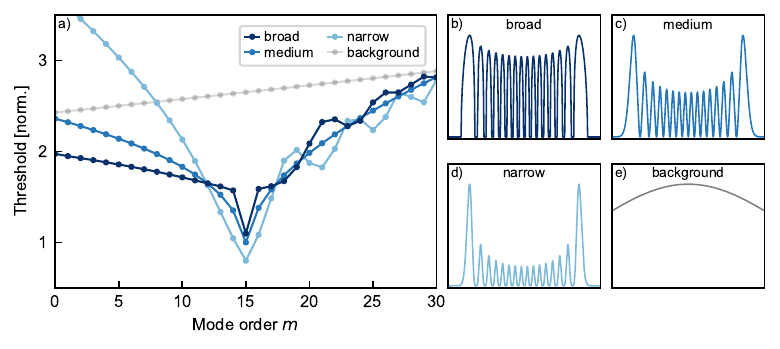}
	\caption{a) Calculated modal thresholds for the used pump patterns to excite the $\text{HG}_{15,0}$ mode shown in b)-e). Though only defined for integer $m$, marked by dots, lines are added to guide the eye. The narrow pump pattern discriminates the target mode well against low-order modes, whereas the broad pump pattern makes the target mode's threshold more distinct compared to neighboring modes. In contrast, the mode-unspecific background beam does not lead to a distinctly reduced threshold of any single mode. All thresholds are normalized to the threshold value of the medium pump pattern at $m=15$. All pump distributions are normalized to peak intensity for better visibility.}
	\label{fig:hg15_thresholds}
\end{figure}

% Pump patterns lead to characteristic threshold curves
For example, all three investigated DMD pump patterns lead to the lowest threshold for the target $\text{HG}_{15,0}$ mode (see Fig.~\ref{fig:hg15_thresholds}), because of the similarity between the spatial intensity distributions of pump beam and resonator mode.
However, the pump patterns vary in how specific they are to the target mode of interest.
The broad pump pattern leads to a steep dip in the threshold curve, i.e., the threshold of the target mode is well separated from that of the neighboring modes. 
The thresholds of low-order modes, however, are also low when compared to the other patterns.
In contrast, the narrow pump pattern shows increased low-order mode thresholds, i.e., it more strongly hinders oscillation of these modes, but the distinction between thresholds of target and neighboring modes is reduced, possibly leading to their excitation along with the target mode.
These characteristics were not only seen for the exemplary $\text{HG}_{15,0}$ mode, but for other mode orders as well (see Fig.~S1 in Supplement~1 for the threshold curves of the other mode orders).
% Modulation in threshold curves
In the cases of the broad and narrow pump patterns, the threshold curves were seen to be modulated for mode orders greater than the target mode order $m=15$ (see Fig.~\ref{fig:hg15_thresholds}a), which, generally speaking, seems plausible given the periodicity of HG modes and how their spatial structure changes with mode order.
In contrast, the medium pump pattern is exactly equal to the intensity distribution of the mode and does not show this modulation.
Hence, there appears to exist a property of the HG mode set that yields smoother overlap integrals and therefore threshold curves when both mode and pump distribution are exactly equal to a HG mode (similar to observations made in \cite{Chen1997}).

% Threshold background beam
The background beam does not lead to any distinct mode with reduced threshold because of its mode-unspecific shape. 
Instead, the modal thresholds only slightly increase with mode order, e.g., the threshold of the fundamental mode is \qty{16}{\percent} lower than that of the $\text{HG}_{30,0}$ mode.
This dependence is due to the broad but still Gaussian intensity distribution of the background beam; a homogeneous flat-top beam results in equal calculated thresholds for all modes.
Therefore, it stands to reason that with increasing background beam power modes of low orders are excited first.

% Combined pump beam
Finally, the combined pump distribution is the power-weighted average of the DMD beam and the background beam. 
Accordingly, the threshold curves depicted in Fig.~\ref{fig:hg15_thresholds} represent the limiting cases of a continuum between mode-specific, when pumping only with the DMD beam, and mode-unspecific, when pumping only with the background beam (see Fig.~S2 in Supplement~1 for an exemplary visualization of that continuum).
In general, the combined pump beam maintains the discussed characteristics, i.e., it exhibits the lowest threshold for the target mode and suppresses other modes to an adjustable degree, depending on the used DMD pump pattern and beam power ratio.
This approach allows to alter the effective pump distribution simply by changing the power ratio between the two pump beams.

\section{Experimental results}
\subsection{From selective single-mode excitation to multi-mode operation}
The combination of a mode-specific pump beam shaped by a DMD and a mode-unspecific background beam was used to excite $\text{HG}_{m,0}$ modes.
% Modal decomposition
To identify the excited transverse modes, a modal decomposition of the output beam based on its intensity distribution captured with a camera (Cam, see Fig.~\ref{fig:setup}) was performed.
Because of the small vertical extents of the pump beams, only modes of the form $\text{HG}_{m,0}$ were excited, such that the intensity distribution could be projected onto the horizontal axis for reduced computational complexity.
Then, an L-BFGS-B optimization algorithm~\cite{Byrd1995, Zhu1997, Morales2011} was used to determine the relative power $W_m$ of each transverse mode by minimizing the least-squares error between measured and numerically reconstructed one-dimensional intensity distribution~\cite{Bruening2013}.

\begin{figure}[!ht]
	\centering
	\includegraphics{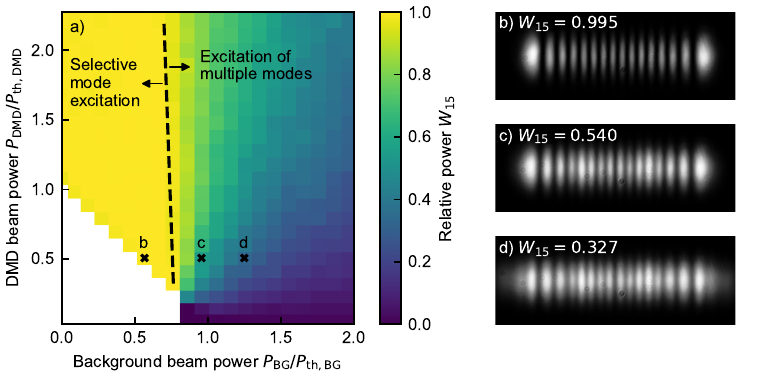}
	\caption{a) Power of the target mode $\text{HG}_{15,0}$ as a fraction of the output beam power (obtained from the modal decomposition) in the case of the medium DMD pump pattern. For low background beam power only oscillation in the target mode was observed, whereas with high background beam power multiple other modes were excited as well. The dashed line indicates the approximately linear border, where the relative power of the target mode was $W_{15}=\num{0.98}$. In the white region (lower left) no laser oscillation occurred. b), c), d) Output beam profiles for the indicated combinations of DMD and background beam powers.}
	\label{fig:hg15opt_thresholdmap}
\end{figure}

% Example HG15
Depending on the ratio between mode-specific pump power of the DMD beam $P_\text{DMD}$ and mode-unspecific pump power of the background beam $P_\text{BG}$, laser oscillation from transverse single-mode operation to oscillation of multiple transverse modes was observed, which is shown in Fig.~\ref{fig:hg15opt_thresholdmap} as an example for the $\text{HG}_{15,0}$ mode with the medium pump pattern (compare Fig.~\ref{fig:pump_patterns}b and e).
The powers $P_\text{DMD}$ and $P_\text{BG}$ of the two pump beams were normalized by their individual thresholds $P_\text{th,DMD}$ and $P_\text{th,BG}$, i.e., by the pump powers necessary to start laser oscillation when only pumping with the DMD beam or the background beam, respectively.
For completeness, the threshold $P_\text{th,DMD}$ increased with mode order and pump pattern (from narrow and medium to broad) due to the different spatial overlap between mode and pump beam (see Eq.~(\ref{eq:modevolume})), but this explicit dependence was omitted from the axis label in Fig.~\ref{fig:hg15opt_thresholdmap} for brevity.

When pumping the gain medium with the mode-specific DMD beam alone ($P_\mathrm{BG}/P_\mathrm{th,BG}=0$) only the target mode was excited, which is evident by the close to unity relative power $W_{15}$.
Adding the mode-unspecific background beam reduced the necessary power of the DMD beam below its initial threshold value while maintaining transverse single-mode operation (see Fig.~\ref{fig:hg15opt_thresholdmap}b).
In this example, the DMD beam power -- and therefore also the power incident on the DMD -- could be reduced to \qty{28}{\percent} of the initial threshold value, which was determined by a separate measurement with greater resolution than in Fig.~\ref{fig:hg15opt_thresholdmap}.
Increasing the background beam power further, eventually led to excitation of other modes (see Fig.~\ref{fig:hg15opt_thresholdmap}c and d), because the background beam provided gain for a wide range of transverse modes.
For visualization, a relative power of $W_{15}=\num{0.98}$ was chosen as the border between single- and multi-mode laser oscillation (dashed line in Fig.~\ref{fig:hg15opt_thresholdmap}).
It is noteworthy that this border was almost independent of the DMD beam power and was close to the threshold power of the background beam.
In other words, once the background beam reached the threshold of any mode, this mode was excited as well, independent of the mode excited by the DMD beam.
This indicates only little interaction between the target mode and other undesired modes despite the homogeneous line broadening of Nd:YVO$_4$, which allows for gain competition between modes and thus favors longitudinal single-mode operation when having only a single transverse mode in operation~\cite{Koechner2006}.
However, with more than one transverse mode gain competition can only occur where the modes' spatial intensity distributions have significant overlap, so that this interaction is essentially limited to modes of similar mode order.
Therefore, when gain is spatially spread out, as by the background beam, there is only a weak dependence between the oscillations of different transverse modes, especially if their mode orders differ substantially.

% Reductions for other pump patterns (and mode orders)
The reduction in effective threshold due to the additional background beam was not only observed for the exemplary $\text{HG}_{15,0}$ mode with the medium DMD pump pattern, but also for the other investigated mode orders and DMD pump patterns (see Figs.~S3 and S4 in Supplement~1 for the corresponding measurements).
The mode-specific DMD pump beam power could be reduced more strongly for lower mode orders, due to their greater overlap with the background beam, and narrow pump patterns, due to the better discrimination against unwanted low-order modes.
To give the two extreme cases as examples, the necessary DMD pump beam power for selective mode excitation was reduced down to \qty{20}{\percent} for $m=5$ with the narrow pump pattern and down to \qty{36}{\percent} for $m=25$ with the broad pump pattern compared to their respective DMD-beam-only thresholds.

% Demonstration for resonator with higher losses
To summarize this section, selective mode excitation by pump beam combining was shown, which allowed the mode-specific pump power to be reduced to between \qty{20}{\percent} and \qty{36}{\percent} of the initial threshold, depending on the mode order and pump pattern.
In other words, the mode-specific pump power can be reduced by a factor of three to five by adding the mode-unspecific background beam.
This, for example, makes possible the excitation in a resonator configuration with three to five times higher losses, which was experimentally verified by replacing the outcoupling mirror with \qty{0.5}{\percent} transmission by one with \qty{2}{\percent} transmission.
Then, due to the increased losses, the DMD beam alone did not allow to excite the $\text{HG}_{30,0}$ mode, even with the broad pump pattern which enabled the highest pump powers through the greater number of 'on' DMD micromirrors.
As expected, with the added background beam, excitation of this mode was again made possible, because of the effective reduction of necessary DMD power.
Alternatively, the additional gain could also be supplied by adapting the pump pattern on the DMD instead.
Here, the binarization threshold $\eta$ of the broad pump pattern was decreased from \num{0.11} to \num{0.06}, expanding the mode-like area of switched 'on' micromirrors in both spatial directions, such that the effective pump power was increased by \qty{25}{\percent} for the same power incident on the DMD.
This can be understood as the broader -- and therefore more efficient -- pump pattern taking on the background beam's function of providing more -- but less mode-specific -- gain, which also enabled excitation of the $\text{HG}_{30,0}$ mode.

\subsection{Modal analysis of the multi-mode operation}
As shown in the previous section, with increased background beam power, the relative power of the target mode is reduced as other modes start to oscillate.
Here, the results of the modal decomposition are analyzed in regard to which modes are excited in addition to the desired mode and how they relate to the used pump pattern.

\begin{figure}[!ht]
	\centering
	\includegraphics{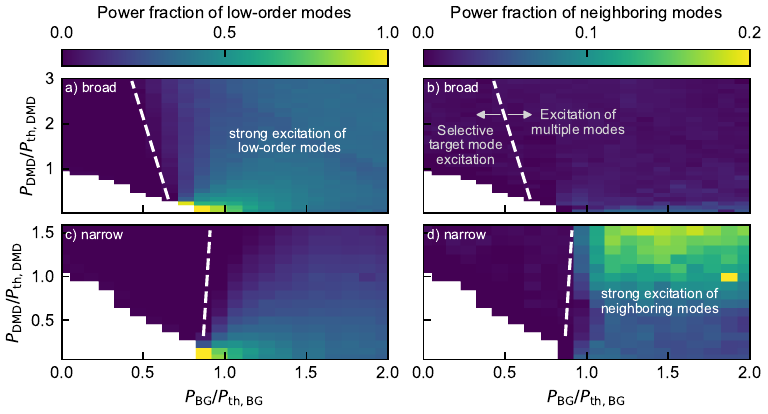}
	\caption{Excitation of unwanted modes besides the $\text{HG}_{15,0}$ target mode. a), c) Low-order modes, i.e., $\text{HG}_{m,0}$ with $0\leq m \leq 13$, are predominantly excited by the broad pump pattern. b), d) Neighboring modes, i.e., $\text{HG}_{14,0}$ and $\text{HG}_{16,0}$, are more strongly excited for the narrow pump pattern. The dashed lines indicate where the relative power of the target mode was $W_{15}=\num{0.98}$ and represent the approximately linear borders between selective mode excitation on the left and multi-mode operation on the right, as indicated in b).}
	\label{fig:hg15_lower_neighboring}
\end{figure}

With the broad DMD pump pattern, low-order modes were predominantly excited when transitioning from single- to multi-mode laser operation (see Fig.~\ref{fig:hg15_lower_neighboring}a), decreasing the relative power of the target mode (compare Fig.~\ref{fig:hg15opt_thresholdmap}).
This is clearly seen as the border between selective excitation and multi-mode operation (dashed lines in Fig.~\ref{fig:hg15_lower_neighboring}) coincided with the onset of oscillation of low-order modes, identifying them as the additionally excited ones.
In comparison, neighboring modes were not excited to significant amounts (see Fig.~\ref{fig:hg15_lower_neighboring}b).
For the narrow DMD pump pattern the opposite was observed: here, the neighboring modes were more strongly excited when the background beam power was increased and multi-mode lasing started (see Fig.~\ref{fig:hg15_lower_neighboring}d).

These experimental observations can be attributed to the characteristic modal thresholds of the broad and narrow pump patterns (see Fig.~\ref{fig:hg15_thresholds} for the calculated thresholds): the broad pump pattern features a very distinct threshold of the target mode compared to its neighbors, but comparatively low thresholds of low-order modes.
Thus, as observed in the experiment, with the broad pump pattern low-order modes are predominantly excited.
In contrast, the narrow pump pattern offers a greater discrimination against low-order modes, at the cost of similar thresholds between target and neighboring modes, resulting in their excitation along with the target mode.
Thus, to a certain degree, the experimentally observed multi-mode laser operation can be derived from simple spatial overlap calculations (see Sec.~\ref{subsec:thresholds}), where the modes are assumed to be independent, although in reality they are at least partially coupled in the laser medium through competition for the limited, spatially distributed gain provided by the pump.

% Thought experiment: pumping with inner and outer lobes
The differences between broad and narrow pump patterns can be understood by considering two extreme cases: pumping with a distribution only consisting of the two outermost lobes of a $\text{HG}_{m,0}$ mode or alternatively pumping with the inverse pattern only consisting of the inner lobes (see Fig.~S5 in Supplement~1 for a visualization of these pump patterns and their thresholds).
The outermost lobes of HG modes are very similar between neighboring modes ($m \pm 1$), because they change only little in position with mode order $m$~\cite{Siegman1986}.
Therefore, all pump beams that have a great fraction of their power off-center, result in similar thresholds for the target mode and its neighbors.
In contrast, the inner lobes are very characteristic for a certain mode compared to its neighbors, because they are 'out-of-phase' with each other, leading to a distinct reduction of the lasing threshold. 
Consequently, pump beams that have a great fraction of their power in the inner lobe structure discriminate well between neighboring modes (broad pump pattern).
The opposite is true for pump beams with a great fraction of their power off-center (narrow pump pattern): they discriminate well against low-order modes, because they provide only little gain near the optical axis where low-order modes have significant intensity.
The medium pump pattern can be seen as a compromise between these two extremes, with moderate suppression of low-order modes and moderate similarity of neighboring mode thresholds.

% Stress general validity and trade-off
We want to emphasize that these properties result from the intensity distributions of the HG mode family and are therefore not specific to the presented experimental implementation, but applicable to all mode-like spatial gain-shaping approaches.
In particular, it can be said that mode-like pumping requires a trade-off between favoring the excitation of a target mode through lowering its threshold and suppressing the excitation of other modes.

\subsection{Exploration of output power scaling}
So far, the background beam served as a tool to purposefully reduce the modal-specificity of a combined pump beam.
As laid out in Sec.~\ref{sec:setup}, the optical power shaped by a DMD is limited by the device's damage threshold, so that after a certain point the pump power applied to the gain medium cannot be increased further by simply raising the power incident on the DMD.
Therefore, we investigated whether it is possible to increase the output power of a given mode by providing more gain through the addition of the background beam.
Again, the $\text{HG}_{15,0}$ mode is given as a representative example.

\begin{figure}[!ht]
	\centering
	\includegraphics{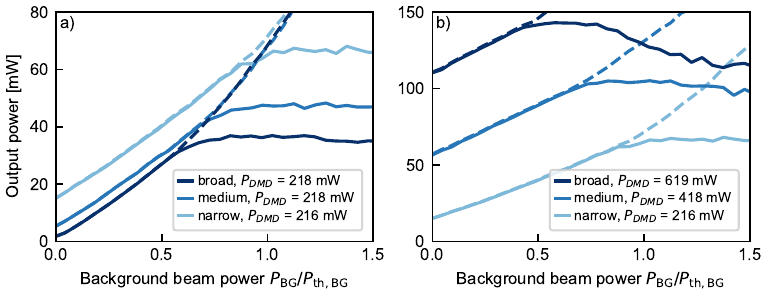}
	\caption{Output power scaling of the target $\text{HG}_{15,0}$ mode by addition of the background beam. Solid lines: power of target mode. Dashed lines: total power of output beam. $P_\text{DMD}$ denotes the pump power incident on the gain medium after shaping with the DMD. a) Equal DMD beam power (after shaping) incident on the gain medium. b) Equal pump power incident on the DMD, leading to different DMD beam powers incident on the active medium due to the different number of 'on' micromirrors of the pump patterns.}
	\label{fig:hg15_output_power}
\end{figure}

% same power after DMD
For the same pump power of the DMD beam incident on the gain medium ($P_\text{DMD}\approx\qty{218}{\mW}$), the overall power of the output beam (dashed lines) and the power of the target mode (solid lines) were determined for pumping with the three distinct DMD patterns (see Fig.~\ref{fig:hg15_output_power}a).
When pumping with the DMD beam alone ($P_\mathrm{BG}/P_\mathrm{th,BG}=0$) only the target mode was excited.
The output power was highest in the case of the narrow pump pattern, as a result of the low threshold value compared to the other pump patterns (compare Fig.~\ref{fig:hg15_thresholds}). 
With increasing background beam power, the output power increased as well: at first in transverse single-mode operation, but eventually other modes were also excited (visually apparent as the splitting of solid and dashed lines), after which the power of the target mode stayed nearly constant.
This again speaks to the fact that there is little interaction between transverse modes and that the background beam predominantly excites modes other than the target mode.
Similar to the presented threshold reduction, the narrow pump pattern could tolerate more mode-unspecific gain of the background beam before other -- mainly neighboring -- modes were excited, resulting in twice the output power (\qty{63}{\mW} compared to \qty{32}{\mW}) of the broad pump pattern with the same level of selective mode excitation ($W_{15}>\num{0.98}$). 

% same power before DMD
However, when considering the damage threshold of the DMD, the power incident on the device is obviously relevant.
With the same power incident on the DMD, (see Fig.~\ref{fig:hg15_output_power}b), the reverse trend was observed: here, the broad pump pattern resulted in the highest output power. 
This was because of the higher achievable pump powers (compare Sec.~\ref{subsec:dmd_shaping}): the broad pump pattern utilized more DMD micromirrors in their 'on'-position, so that for the same power incident on the DMD almost three times as much pump light (\qty{619}{\mW}) was actually shaped and redirected towards the gain medium compared to the narrow pump pattern (\qty{216}{\mW}).
When extrapolating this trend, one arrives at the recommendation to utilize as much of the power incident on the DMD by making the pump pattern as broad as possible for maximum shaping efficiency and using no background beam at all, when aiming for output power scaling.

% outlook different shaping devices
The -- in hindsight obvious -- recommendation of maximizing the mode-specific gain, can also be extended to other beam shaping devices.
Liquid-crystal-based spatial light modulators are similar to DMDs in their damage thresholds and also allow for flexible adaption of the pump pattern. 
In contrast, fixed beam shaping devices, such as DOEs or photolitographic masks, only produce a single beam shape, but are usually not limited by their damage thresholds, so that the power incident on the shaping device can be increased for output power scaling~\cite{Liu2004, Anoikin2015, Hilton2015}.
Therefore, also in these cases, the efforts of adding a second pump beam and implementing a beam combining scheme can be avoided, because higher output power through additional gain is more easily realized by either adapting the pump pattern or simply increasing the pump power.

\section{Can mode-like pumping be improved upon?}
Usually, as done in this paper (see Sec.~\ref{subsec:dmd_shaping}) and in other gain shaping setups~\cite{Schepers2019, Zhang2021}, pump distributions are derived from the intensity profile of the target mode by some kind of function conceived by a human, e.g., being exactly equal to the mode's intensity or slightly adapted through a binarization threshold.
But we pondered the question of whether such mode-like pump distributions, that are visually similar to the target mode's intensity profile, were actually the best to selectively excite the target mode or if they could be improved.
Maybe there are other, possibly non-intuitive and complex, spatial distributions which more exclusively provide gain only for a single mode and thus hinder excitation of other transverse modes.

We addressed the above question by using a numeric optimization approach to develop initial random spatial intensity distributions into potentially novel pump distributions.
The previous experiments showed that a pump pattern is well suited for selective excitation if it yields the lowest lasing threshold for a target mode and, ideally also, distinctively higher thresholds for all other modes. 
Hence, to explore the influence of these objectives, pump patterns were optimized based on two figures of merit: one, $f_1$, that minimizes the target mode threshold and one, $f_2$, that minimizes the target mode threshold and simultaneously maximizes the separation in threshold value from any other mode. 
This optimization approach has the potential to come up with previously unthought-of pump distributions, because they are developed numerically based on their desired properties (modal thresholds), instead of being derived directly from the intensity of the target mode.

In a first step, that pump pattern was determined which minimized the threshold of the target mode. 
The intensity of the pump pattern $I_\text{pump}(x)$ at each spatial position $x$ was given as an adjustable parameter to an optimization algorithm, while using the threshold $P_\text{th}$ of the target mode $m_0$ as the figure of merit $f_1$ to minimize:
\begin{equation}
	f_1 = P_\mathrm{th}(m_0).
\end{equation}
Starting with a random initial pump distribution $I_\text{pump}(x)$, one after another, a differential change was made to the pump intensity at each discrete spatial position, after which the modal thresholds were calculated based on Eq.~(\ref{eq:modevolume}) and the influence (namely, first- and second-order partial derivatives) on the figure of merit was evaluated.
Then, the pump distribution was updated accordingly to iteratively minimize the figure of merit.
The numeric optimization was performed by an L-BFGS-B algorithm, which was chosen for two reasons~\cite{Byrd1995, Zhu1997, Morales2011}. 
First, it allows to restrain parameters within boundaries, i.e., $I_\text{pump}(x)>0$ to prevent unphysical negative intensity values.
Second, it is capable of handling many parameters, which were required for sufficient spatial resolution, as each parameter represents the pump intensity at one spatial point.
Here, the pump distributions were spatially sampled at \num{301} points (later, $61^2=\num{3172}$ points were used for two-dimensional pump patterns for $\text{HG}_{m,n}$ modes).

\begin{figure}
	\centering
	\includegraphics{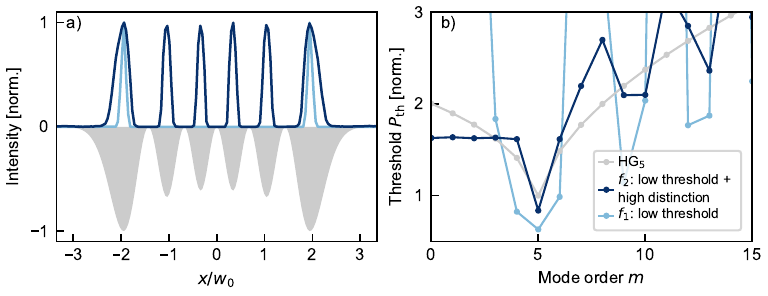}
	\caption{a) Numerically optimized pump patterns (lines) compared to the intensity distribution of the target mode ($m_0=5$, gray shading, flipped down). For better visibility of spatial features the  $\text{HG}_{5,0}$ mode was chosen as a representative example and the pump distributions were individually normalized to maximum intensity. b) Corresponding threshold curves; though only defined for integer $m$, marked by dots, lines are added to guide the eye.}
	\label{fig:optimized1d}
\end{figure}

The resulting pump pattern found through optimization of $f_1$ consists of two intensity maxima at the mode's outermost peaks (see light blue curve in Fig.~\ref{fig:optimized1d}a) and thus resembles the familiar pump distribution in a symmetric off-axis pumping scheme.
For comparison, the lasing thresholds for pumping with a distribution equal to the intensity of the target mode (see gray shading in Fig.~\ref{fig:optimized1d}a) and the corresponding modal thresholds are also shown (see gray curve in Fig.~\ref{fig:optimized1d}b).
This mode-identical pump distribution serves as a reference and all thresholds are normalized to its value at the target mode (notice how the gray curve in Fig.~\ref{fig:optimized1d}b defines $P_\mathrm{th}(m_0)=1$).
As expected, the optimized pump distribution has the lowest lasing threshold for the target mode, with the threshold value being \qty{37}{\percent} lower than for the mode-identical pump distribution.
However, the thresholds of neighboring modes are more similar to that of the target mode.

Then, to achieve a minimal target threshold and simultaneously maximal separation from other modes, the figure of merit was extended to
\begin{equation}
	f_2 = \alpha P_\mathrm{th}(m_0) - \min[P_\mathrm{th}(m \neq m_0)], \label{eq:objective_balanced}
\end{equation}
where the first term represents the minimization of the target mode threshold and the additional second term is responsible for maximizing threshold separation.
Consequently, during the optimization process, the figure of merit $f_2$ is reduced when the threshold of the target mode $m_0$ is decreased and the lowest threshold (obtained via the minimum operation $\min[\cdot]$) of all the other modes $m \neq m_0$ is increased.
The two objectives are weighted by a factor $\alpha$ relative to each other, and a value of $\alpha=\num{2}$ was empirically found to lead to balanced results that embody both a low and distinct target mode threshold.
The optimized pump pattern (see dark blue curve in Fig.~\ref{fig:optimized1d}a) roughly retains the shape of the target mode, but differs insofar as the pump distribution is narrower than the mode profile and the peak intensities are approximately of equal height -- in contrast to the HG mode profile with less intense inner lobes.
Compared to the gain distribution identical to the mode's intensity, the optimized pump pattern shows a \qty{16}{\percent} lower threshold due to its narrower features and a more distinct threshold (\qty{37}{\percent} greater separation to other modes' thresholds) due to the more center-heavy pump distribution (see Fig.~\ref{fig:optimized1d}b).
However, the thresholds of low-order modes are reduced to the same level as the neighboring modes, because of the greater pump power near the optical axis, leading to a similar threshold curve as the 'broad' pump pattern presented earlier, which also shows threshold values that are nearly constant for low-order modes and modulated for high-order modes (see Fig.~\ref{fig:hg15_thresholds}).

Both optimized pump distributions, relating to $f_1$ and $f_2$, illustrate the previously discussed qualitative relationships deduced from the experimental data (see Fig.~\ref{fig:hg15_lower_neighboring}): pump patterns with a great fraction of their power away from the optical axis lead to low thresholds of the target mode and its neighbors.
In contrast, center-heavy pump distributions improve the distinction between the target mode and its neighbors.
Entering the optimization, the pump distributions were initialized as noise and despite their random initial states, the pump distributions shown in Fig.~\ref{fig:optimized1d} repeatedly emerged in high similarity (median correlation greater than \num{0.95} in \num{50} optimization runs), indicating a global optimum.
The generality of the optimized distributions was further emphasized as the characteristics of the so optimized pump patterns were found not only for the shown $\text{HG}_{5,0}$ mode, but also for other $\text{HG}_{m,0}$ modes and even two-dimensional $\text{HG}_{m,n}$ modes (see Figs.~S6 and S7 in Supplement~1 showing the other optimized pump patterns).
Also for these modes, the optimized pump patterns were in principle similar to the modes' intensity distributions but had overall narrower and more intense features in the center of the pump patterns.
Thus, to come back to the original question of whether a possibly complex non-mode-like pump distribution could improve performance in selective excitation, it appears that a distribution similar -- although not identical -- to the target mode's intensity is more beneficial.

\section{Conclusion}
We explored the excitation of transverse laser modes by spatially structured pump light, with special attention given to the influence of the similarity between pump and modal intensity distributions.
To study the transition from selective excitation of only a single mode to lasing of multiple transverse modes, in the experiments, pump distributions were deliberately made less similar to a mode by superposing a mode-specific pump beam with a mode-unspecific background beam.

Analysis of the usually unwanted multi-mode excitation revealed characteristic properties of pump distributions adapted to Hermite-Gaussian $\text{HG}_{m,0}$ modes: a center-heavy pump distribution at first distinctly excites the target mode and eventually low-order modes, whereas an eccentric pump distribution reduces the lasing threshold at the expense of distinction to the neighboring modes. 
These trends originate in the spatial overlap of the intensity profiles of HG modes and can therefore also be applied to other gain shaping setups, e.g., using different gain media or beam shaping devices, such as liquid-crystal based spatial light modulators and diffractive optical elements.
For the same reason, these characteristics of spatial gain distributions are also valid for two-dimensional $\text{HG}_{m,n}$ modes, and similarly also apply to Laguerre-Gaussian $\text{LG}_{l,p}$ modes as they exhibit comparable spatial scaling with mode order, after accounting for the difference between Cartesian and radial symmetry.
In all cases, selective excitation of a single mode can potentially be improved by applying modified gain distributions that take into account the presented spatial overlap characteristics.

Finally, a numeric optimization was employed to explore whether non-intuitive and possibly complex pump distributions could improve selective excitation compared to mode-like pumping.
The optimizations were driven by two figures of merit, each derived from modal threshold calculations.
The resulting pump distributions were not equal to the target modes' intensity profiles, but they could still be described as mode-like, because they maintained the general structure of the modes and showed only minor changes, e.g., overall narrower features or more power close to the center of the distribution.
Surely, with different figures of merit, other optimal pump distributions would emerge, but the presented results make it seem unlikely that these would be qualitatively different and not be a recognizable variant of the target mode.
These findings confirm the mode-like pumping approach in existing gain shaping setups and underline that the characteristic spatial intensity distributions of transverse modes are well suited to distinctly excite single transverse modes.

\begin{backmatter}
\bmsection{Funding}
Open Access Publication Fund of the University of Münster.
\bmsection{Acknowledgment}
We acknowledge support from the Open Access Publication Fund of the University of Münster.
\bmsection{Disclosures}
The authors declare no conflicts of interest.

\bmsection{Data Availability Statement}
Data underlying the results presented in this paper are not publicly available at this time but may be obtained from the authors upon reasonable request.

\bmsection{Supplemental document}
See Supplement 1 for supporting content. 

\end{backmatter}

%%%%%%%%%%%%%%%%%%%%%%% References %%%%%%%%%%%%%%%%%%%%%%%%%
\bibliography{bib}

\end{document}

\maketitle
\cleardoublepage

\section{Calculated thresholds for mode-like pumping}
\begin{figure}[h]
	\centering
	\includegraphics{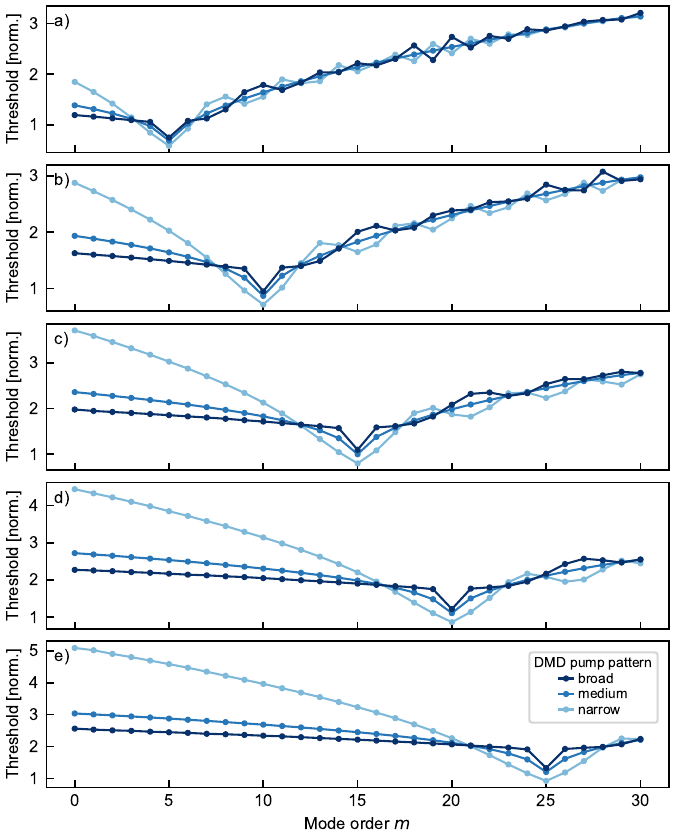}
	\caption{Mode-like pumping reduces the lasing threshold for the target mode. The pump patterns are based on the intensity distributions of the $\text{HG}_{m,0}$ modes with a)-e) $m=\numlist{5;10;15;20;25}$, respectively. The threshold values are all normalized by the threshold value for $m=15$ with the medium pump pattern.}
	\label{fig:tresholds_calc}
\end{figure}

Fig.~3a in the main document shows the threshold curve for the pump patterns designed to excite the $\text{HG}_{15,0}$ mode.
Here, the threshold characteristics for the other investigated mode orders are shown for completeness (see Fig.~\ref{fig:tresholds_calc}).
By adapting the pump distribution to the spatial intensity distribution of a certain mode, the threshold of that target mode is reduced compared to other modes, allowing for its excitation.

The pump patterns (broad, medium, narrow; for definition and examples, see Sec.~3 in the main document) differ in how specific they are to a target mode: for the broad pump pattern the threshold of the target mode is very distinct from that of neighboring modes. 
In contrast, the narrow pump pattern leads to a greater difference between thresholds of target and low-order modes.
These qualitative properties are shared between all mode orders.

\cleardoublepage
\section{Intensity and threshold of combined pump distribution}
\begin{figure}[h]
	\centering
	\includegraphics[width=\textwidth]{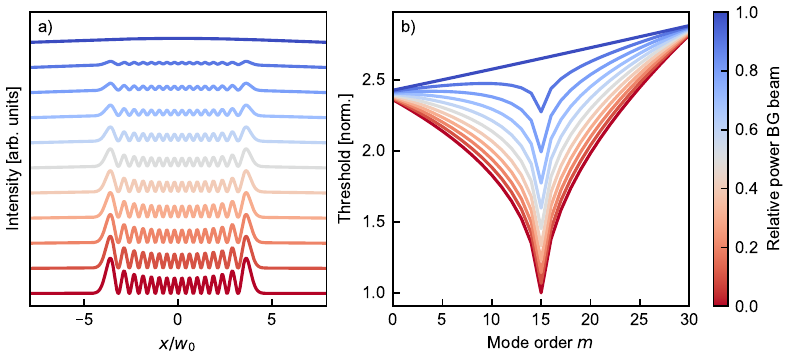}
	\caption{Varying the powers of a mode-specific DMD pump beam and a mode-unspecific background (BG) beam opens up a continuum of a) effective pump distributions (vertically offset for better visibility) and b) corresponding modal thresholds. Both metrics show the characteristic range from mode-specific pumping when only using the DMD beam (red, BG beam power $=0$) to mode-unspecific pumping when only using the BG beam (blue, BG beam power $=1$). The $\text{HG}_{15,0}$ mode with the medium pump pattern is used as a representative example.}
	\label{fig:thresholds_average}
\end{figure}

\cleardoublepage
\section{Modal decomposition of transverse multi-mode lasing}
\begin{figure}[h]
	\centering
	\includegraphics{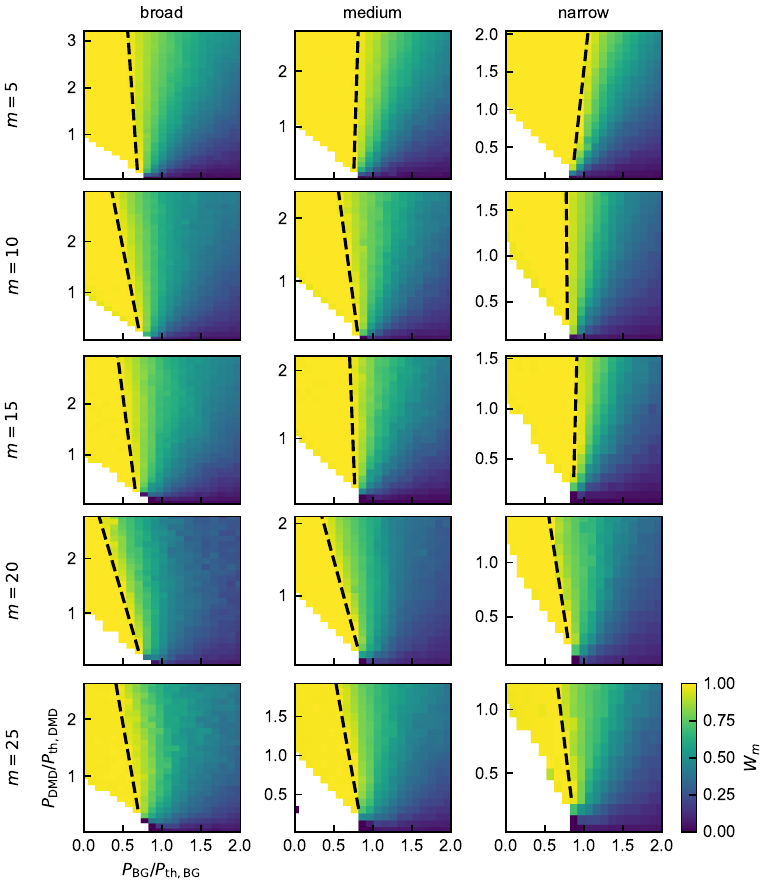}
	\caption{Measured level of selective mode excitation as a function of DMD beam power $P_\text{DMD}$ and background beam power $P_\text{BG}$, normalized by their respective thresholds, for different $\text{HG}_{m,0}$ target modes and DMD pump patterns (broad, medium, narrow). The dashed lines indicate the border between selective mode excitation and transverse multi-mode operation with $W_m=0.98$.}
	\label{fig:treshold_map}
\end{figure}

Fig.~4a in the main document shows the relative power $W_m$ of the target mode as a fraction of the total laser output power for different DMD and background beam powers in the case of the medium pump pattern and the $\text{HG}_{15,0}$ mode.
Here, the modal purity of the laser output is shown for the other experimentally investigated mode orders and pump patterns for completeness (see Fig.~\ref{fig:treshold_map}). 
Qualitatively, all measurements show the same trend: when using only the DMD -- and thus applying a mode-specific pump -- a normalized pump power of $P_\text{DMD}/P_\text{th,DMD} = 1$ is required to excite the target mode.
With the addition of the mode-unspecific background beam, the necessary DMD power can be reduced below the initial threshold value $P_\text{th,DMD}$ while still only exciting the target mode.
However, for higher background beam powers multiple transverse modes are excited, so that the relative power $W_m$ of the target mode is reduced.
The borders between selective excitation and transverse multi-mode operation were approximately linear (see dashed lines in Fig.~\ref{fig:treshold_map}), which indicates the reduced gain competition between transverse modes due to their different spatial profiles, as discussed in Sec.~2.3 of the main document.

When comparing the broad, medium and narrow pump patterns, the linear border between selective and multi-mode excitation appears to 'shift right' and 'pivot clockwise'.
This means, for the narrow pattern a greater amount of mode-unspecific pump power through the background beam is tolerated before other modes are excited.
As the threshold curves (see Fig.~\ref{fig:tresholds_calc}) suggest and the modal decompositions (see Fig.~5 in the main document) show, narrow pump patterns more strongly prevent the excitation of low-order modes, that are otherwise favored by the background beam, so that these unwanted modes are only excited for comparatively larger powers $P_\text{BG}$ of the background beam.

Looking for a trend across the different mode orders, it seems that low-order modes tolerate more background beam power before other modes are excited, when comparing the border between selective and multi-mode excitation for $m=5$ and $m=25$.
A reason for this could be the -- albeit weak, but still existing -- competition between target mode and other low-order modes for the gain provided by the background beam.
This effect is naturally more pronounced for a low-order target mode ($m=5$). 
However, upon considering the other target mode orders as well, this trend can neither be convincingly confirmed nor refuted.
Measurements with even greater power resolution and more mode orders or simulations taking into account gain competition in the laser medium would allow for a more conclusive assessment, but were deemed beyond the scope of the here presented investigations.

\cleardoublepage
\section{Threshold reduction}
As Fig.~\ref{fig:treshold_map} shows, by applying the additional background beam the power of the DMD pump beam can be reduced while still achieving selective mode excitation ($W_m>\num{0.98}$ even with $P_\text{DMD}/P_\text{th,DMD}<1$).
In a next step, it was determined more precisely how strongly the power of the DMD pump beam could be reduced, by iteratively lowering the DMD beam power while increasing the background beam power until the laser output profile was no longer in single transverse mode operation.
Then, that power of the DMD pump beam is referred to as reduced threshold $P_\text{th,DMD}'$, which is given relative to the corresponding threshold when only using the DMD beam.
For example, a reduced threshold $P_\text{th,DMD}' = \num{0.25}$ means that the power of the DMD pump beam can be reduced to $\num{0.25}\times P_\text{th,DMD}$, i.e., the target mode is selectively excited even though the DMD pump beam power is reduced by a factor of \num{4}, because of the additional pumping with the background beam.

\begin{figure}[h]
	\centering
	\includegraphics{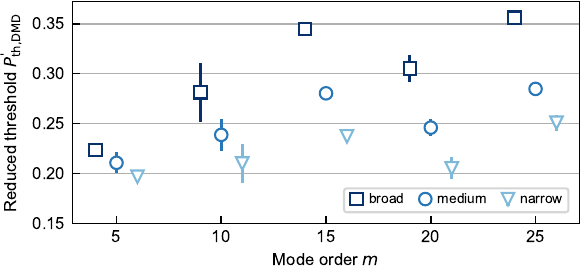}
	\caption{Reduction of necessary DMD power for selective excitation by using the background beam. Small mode orders and narrow pump patterns allow for the greatest reduction of mode-specific pump power $P_\text{DMD}$. Markers of the broad and narrow pump patterns are slightly shifted left and right, respectively, for better distinguishability; error bars indicate the standard deviation of five measurements.}
	\label{fig:threshold_reduction}
\end{figure}

For all investigated mode orders, the DMD beam power could be reduced most strongly in the case of the narrow pump pattern (see Fig.~\ref{fig:threshold_reduction}).
As discussed (see Fig.~\ref{fig:tresholds_calc}), the narrow pump pattern delays the onset of low-order mode oscillation due to its comparatively eccentric power distribution, allowing to further reduce the DMD beam power by adding greater amounts of mode-unspecific background beam power. 

As a second observable trend, the DMD beam power could be reduced furthest for low-order modes, because they are not only pumped by the mode-specific DMD beam but also experience comparatively more gain than high-order modes by the background beam with its wide Gaussian intensity profile.
As a background beam with a uniform intensity, e.g., a top-hat profile, results in equal modal thresholds (following from Eq.~(1) in the main document), it stands to reason that it would also result in equal amounts of threshold reduction over mode order.

Notably, the low reduced thresholds in the case of the $\text{HG}_{20,0}$ mode did not fit the increasing trend with mode order.
The different pump patterns for this mode, however, exhibited the expected trend of highest reduction for the narrow pump pattern.
Additionally, the threshold values were reproducible in five separate measurement runs, which further suggests an underlying systematic or physical cause for these observations.
However, in and of itself the $\text{HG}_{20,0}$ mode does not stand out from the other investigated modes or has features that make it particularly suitable for the investigated laser system.
Other measurements, such as those to determine the initial threshold $P_\text{th,DMD}$, did not show any unexpected differences between mode orders that would motivate further investigations.

\cleardoublepage
\section{Calculated thresholds for pumping only with outer and inner lobes}
\begin{figure}[h]
	\centering
	\includegraphics[width=\textwidth]{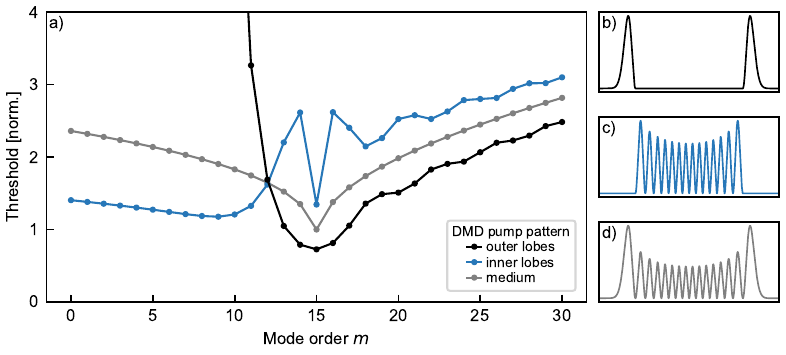}
	\caption{a) Calculated lasing thresholds for pumping the b) outer and c) inner lobes of the $\text{HG}_{15,0}$ mode. The d) 'medium' pump pattern represents a pump pattern adapted to the whole mode profile (see Fig.~2b and e in the main document) and is given for comparison and normalization of the threshold values.}
	\label{fig:thresholds_innerouter}
\end{figure}

When pumping only the outer lobes of a target $\text{HG}_{m,0}$ mode, this target mode benefits most from the so provided gain and hence exhibits the lowest threshold.
Due to the section with zero intensity in the center of this pump distribution, low-order modes do not experience gain, which results in their essentially infinitely high lasing thresholds.
However, such a pump pattern also has significant overlap with the neighboring modes, leading to similar threshold values, because the outermost lobes of the $\text{HG}_{m,0}$ modes change only little with mode order compared to the lobe width.\footnote{The outermost lobe position of a Hermite-Gaussian mode of order $m$ can be approximated as $x_m=\sqrt{m}w$, with $w$ being the fundamental mode radius~\cite[691]{Siegman1986}. Thus, the outermost lobes of neighboring modes are $\Delta x_m = m^{-0.5}\cdot w/2$ apart. As we are not aware of any analytic expression for the widths $W_m$ of the outermost lobes, we determined them numerically from the intensity profile of Hermite-Gaussian modes using the second-order moments definition of width, resulting in an estimation of $W_m=m^{-0.16}\cdot3w_0/4$. Hence, the lobes change only little in position compared to their widths as the ratio $\Delta x_m/W_m=m^{-0.34}\cdot 2/3$ is smaller than unity. Furthermore, $\Delta x_m/W_m$ decreases with mode order $m$, meaning higher mode-orders are more similar to each other, which is consistent with the visual impression upon inspecting the intensity profiles of neighboring modes.}
On the other hand, the inverse pump pattern consisting only of the inner lobes of the target mode leads to very distinct differences in thresholds around the target mode order.
Thus, when only pumping the inner lobes, the threshold of the target mode is well set apart from that of neighboring modes, as the inner lobes of neighboring modes are 'out-of-phase'.
The strong threshold distinction vanishes for modes of other orders, because then the inner lobe structures become again quite similar.
These observations can be generalized to other mode-like pump patterns.
Pump patterns that have a great fraction of their power in the inner lobe structure discriminate well between neighboring modes, whereas pump patterns with a great fraction of their power away from the optical axis discriminate well against low-order modes.

\cleardoublepage
\section{Can mode-like pumping be improved upon?}
\subsection[One-dimensional HGm0 modes]{One-dimensional HG\textsubscript{\textit{m},0} modes}
\begin{figure}[!h]
	\centering
	\includegraphics[width=\textwidth]{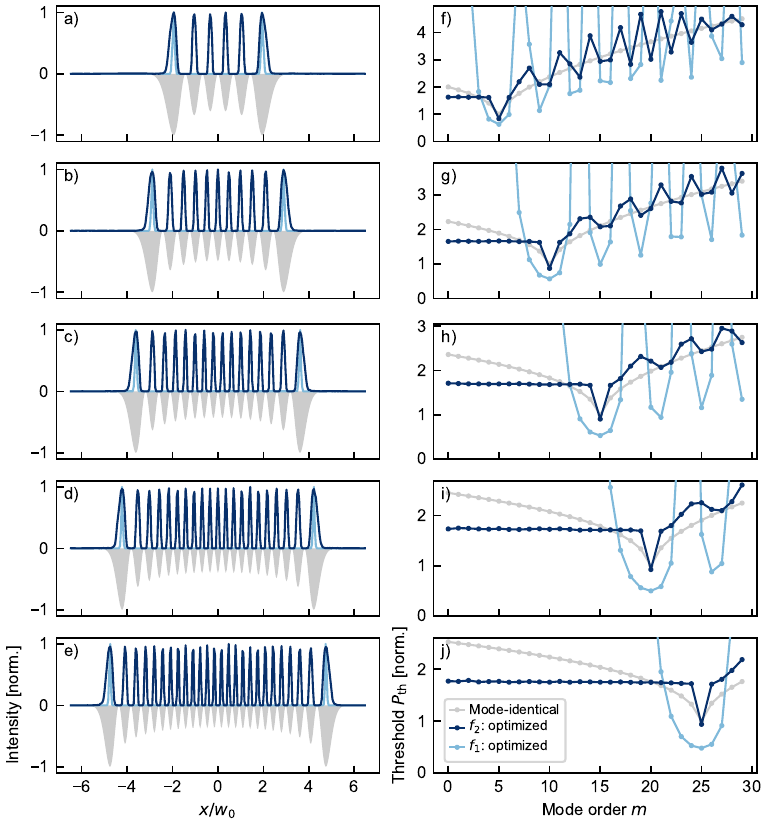}
	\caption{a)-e) Numerically optimized pump patterns for $\text{HG}_{m,0}$ modes with $m=5,\,10,\,15,\,20,\,25$ for two figures of merit, $f_1$ and $f_2$ (light blue and dark blue curves, respectively), as defined in Sec.~5 of the main document. f)-j) The corresponding calculated modal thresholds. Compared to a pump distribution equal to the modal intensity (gray shading, flipped down in a)-e)), the optimized pump distributions feature lower thresholds for the target modes ($f_1$) and also better discriminate against neighboring modes ($f_2$). The intensity profiles are normalized to their respective maximum value, whereas the thresholds are normalized to the minimum of the mode-like pumping approach.}
	\label{fig:hg1d_optimization}
\end{figure}
\cleardoublepage

\subsection[Two-dimensional HGmn modes]{Two-dimensional HG\textsubscript{\textit{m},\textit{n}} modes}
\begin{figure}[!h]
	\centering
	\includegraphics[width=0.94\textwidth]{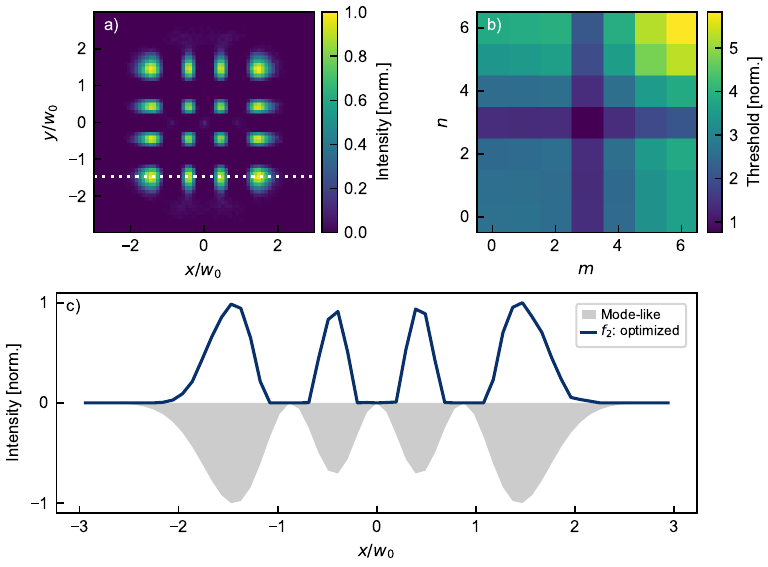}
	\includegraphics[width=0.94\textwidth]{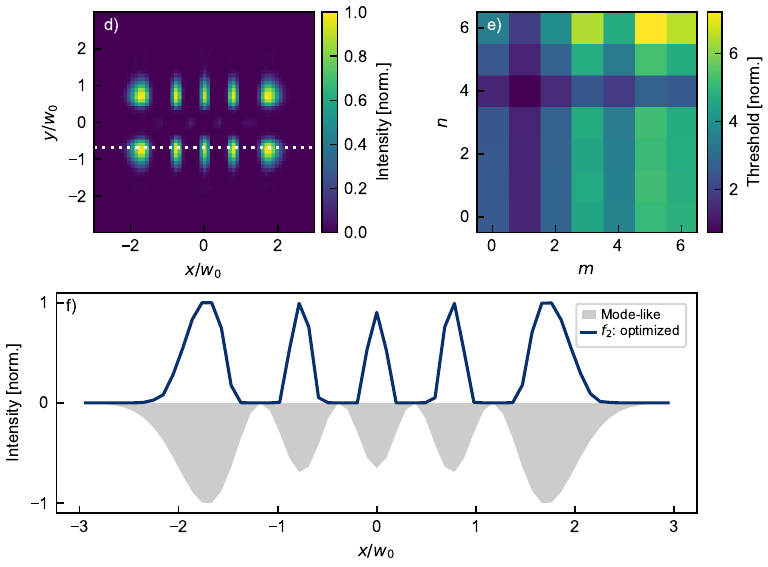}
	\caption{Exemplary optimized pump patterns for $\text{HG}_{m,n}$ modes: a)-c) $\text{HG}_{3,3}$ and d)-f) $\text{HG}_{4,1}$. Similar to the one-dimensional $\text{HG}_{m,0}$ modes (see Fig.~\ref{fig:hg1d_optimization}), the optimized pump patterns exhibit narrower features and their central peaks show a less pronounced intensity decrease, thus concentrating more power near the center compared to the intensity profile of the mode itself. For simplicity, for these two-dimensional modes only the optimization results for the figure of merit $f_2$ are shown (see Sec.~5 in the main document).}
	\label{fig:hg2d_optimization}
\end{figure}

\bibliography{bib}